# Infrared properties of θ-type ET charge-transfer salts:

## θ-ET$_2$RbZn(SCN)$_4$ vs θ-ET$_2$CsZn(SCN)$_4$

**N. L. Wang[1], T. Feng[1], Z. J. Chen[1] and H. Mori[2]**

[1]Institute of Physics, Chinese Academy of Sciences, Box 603, Beijing 100080, PR China

[2]Institute for Solid State Physics, University of Tokyo, Kashiwanoha, Kashiwa-shi, Chiba 277-8581, Japan

___________________________________________________________________________________________________

**Abstract:** Far-infrared reflectance spectra were measured at different temperatures on θ-ET$_2$RbZn(SCN)$_4$ and θ-ET$_2$CsZn(SCN)$_4$, two typical charge-transfer salts in a series of isostructural θ-ET$_2$MM'(SCN)$_4$. The measurements revealed distinct differences in both electronic and vibronic spectra, which provide a clue towards understanding the different metal-insulator transitions of the two salts.

___________________________________________________________________________________________________

The quasi-two-dimensional θ-phase ET-based charge transfer salts θ-ET$_2$X (ET=BEDT-TTF) have recently attracted much attention because they show various unusual physical phenomena. Because of the regular arrangement of donor molecules along c-direction, the θ-ET$_2$X salts with monovalent X$^-$ are expected to have quarter-filled band. On the basis of the band picture, one would expect a metallic behavior in those compounds. However, most of the θ-type compounds undergo metal-insulator (M-I) transitions with decreasing temperature.[1] For θ-(ET)$_2$MM'(SCN)$_4$ (MM'=RbCo, RbZn, CsCo, CsZn), the M-I transition temperatures decrease from 190 K (θ-RbCo, θ-RbZn) to 20 K (θ-CsCo, CsZn). By studying an alloyed system θ-(ET)$_2$(Rb$_{1-x}$Cs$_x$)Zn(SCN)$_4$, Mori et al pointed out that the mechanisms of the metal-insulator transitions for the isostructural θ-RbZn and θ-CsZn salts are different from each other.[2] For θ-RbZn salt, a charge ordering state at low-T associated with a lattice modulation is inferred from various measurements.[3,4] It would be very interesting to investigate the electronic state of θ-CsZn salt at low temperature.

Optical spectroscopy is an important technique for probing the electronic state of a material. Tajima et al[4] performed the first optical reflectance measurement on those θ-type ET salts. They found a Drude-like response at low-frequency for θ-CsZn salt. However, the measurement did not extend to the far-infrared region. In this case, the low-ω extrapolation of reflectivity in the Kramers-Kronig transformation may affect its conductivity spectrum. The present work extends the reflectance measurement to the far-infrared region. The conductivity spectra of θ-CsZn salt were extracted and compared with the θ-RbZn salt.

Figure 1 and figure 2 show the optical conductivity spectra of θ-ET$_2$CsZn(SCN)$_4$ and θ-ET$_2$RbZn(SCN)$_4$ in the polarizations of **E**//a and **E**//c, respectively. At room temperature, the conductivity spectrum of θ-CsZn salt is similar to that of θ-RbZn salt. However, the overall spectral weight is higher in both polarizations, reflecting higher contribution of relatively free carriers. With decreasing temperature, the spectral weight shifts from high-ω to low-ω region. The low-ω conductivity is significantly enhanced. The trend is in sharp contrast to the θ-RbZn salt.

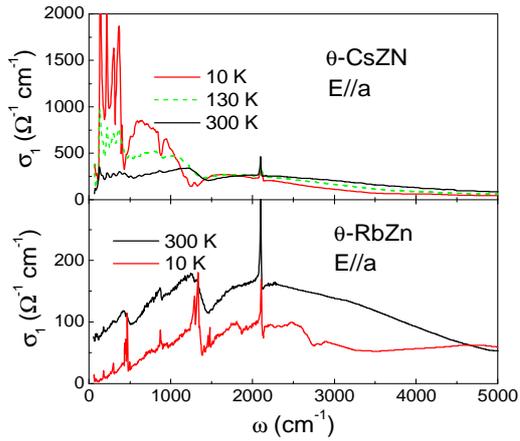

Fig.1. The optical conductivity spectra of θ-ET$_2$CsZn(SCN)$_4$ and θ-ET$_2$RbZn(SCN)$_4$ for **E**//a at several temperatures

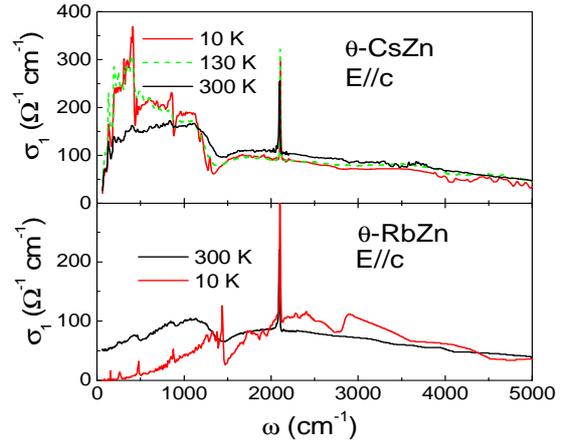

Fig.2. The optical conductivity spectra of θ-ET$_2$CsZn(SCN)$_4$ and θ-ET$_2$RbZn(SCN)$_4$ for **E**//c at several temperatures.

It is noted that the enhancement of low-ω conductivity differs remarkably from an usual Drude-like response. $\sigma_1(\omega)$ decreases sharply at very low-ω, suggesting localized behavior of charge carriers. This is consistent with the insulating behavior observed in the dc resistivity. Nevertheless, a small excitation energy could result in large contribution of charge carriers to the conductivity. This suggests that the charge carriers in this salt locate on the verge of localized and delocalized state. This is very different from the θ-RbZn salt in which an energy gap is formed due to the formation of charge ordering.[5]

Another notable observation is the unusual vibrational spectra. Usually, the vibrational features should be less pronounced with the increase of electronic background. The observed vibrational behavior in θ-CsZn salt seems to be in contrast to this expected behavior. Accompanied by a sharp increase of the electronic background in low frequencies with decreasing T, we find enhanced vibrational structures. In particular, several sharp dips appear in the conductivity spectra. We relate those dips to the intramolecular vibrations being coupled strongly to the charge densities, i.e. an antiresonance of a vibrational spectrum. Those so-called vibronic structures reflect strong electron-molecule vibrational (emv) coupling. Therefore, the antiresonance of emv mode becomes enhanced with an increase of electronic background. In θ-RbZn salt, by contrast, such emv mode becomes ordinary (Lorentzian) vibrational peak since the electronic background is reduced at low temperature.

To conclude, our optical reflectivity measurements revealed distinct differences in both electronic and vibronic spectra. The θ-RbZn salt shows very strong insulating behavior due to the opening of an energy gap at low temperatures, which can be explained as due to the formation of charge-ordering state.[5] By contrast, the θ-CsZn salt shows enhanced electronic spectral weight at low frequencies, although localized electronic behavior was still observed in conductivity spectra. The observation indicates that the θ-CsZn salt locates near the border of metal-insulator transition.